\documentclass[a4paper,11pt]{article}
\pdfoutput=1 
\usepackage{jheppub} 

\usepackage{amssymb}
\usepackage{amsmath}
\usepackage{setspace}
\usepackage{amsfonts}
\usepackage{xcolor}
\usepackage{color}
\usepackage{braket}
\usepackage{bbm}
\usepackage{mathtools}
\usepackage{caption}
\usepackage{subcaption}
\usepackage{enumerate}
\usepackage{bm}
\usepackage[final]{changes}
\definecolor{dolphingreen}{rgb}{0,0.74,0.63}
\definecolor{dolphinorange}{rgb}{0.98,0.53,0.098}
\definecolor{purple}{rgb}{0.4,.2,0.7}
\usepackage{mathrsfs}

\newcommand{\cts}[1]{\ignorespaces}

\usepackage{graphicx}
\graphicspath{{./Figs/}}
\usepackage{cleveref}

\newcommand{\st}[1]{\subsubsection*{\textcolor{blue}{#1}}}
\def\bne{\begin{equation}}
\def\ene{\end{equation}}
\newcommand{\del}{\partial}

\newsavebox{\imagebox}

\title{Island mirages}

\author[a]{Andrew Rolph}

\affiliation[a]{Institute for Theoretical Physics, University of Amsterdam, 1090 GL Amsterdam, The Netherlands}

\emailAdd{andrew.d.rolph@googlemail.com}

\abstract{
We point out a loophole in the proof that the Island Finder conditions~\cite{Bousso2021} are sufficient for the existence of islands. 
We explore examples which satisfy the conditions, but have no islands, which we call island mirages.
We also describe and give resolutions to two new puzzles associated with islands, including a direct tension with the quantum Bousso bound.  
}

\begin{document}


\maketitle
\flushbottom

\section{Introduction}
The black hole information problem (BHIP) is a tension between quantum mechanical unitarity and the insensitivity of Hawking radiation to initial conditions, and has been a persistent and impelling puzzle in quantum gravity since its conception~\cite{Hawking1975,Hawking1976}. The Page curve, the von Neumann entropy of emitted radiation as a function of time, is one sharp diagnostic of unitarity in black hole evaporation~\cite{Page1993}. Deriving a Page curve consistent with unitarity, one that ends at zero von Neumann entropy for a black hole formed from collapse of matter in a pure state, is an important step in resolving the BHIP, though it is not the whole story.  

Much of the recent progress in deriving a Page curve consistent with unitarity has been in developing the techniques to calculate the entropy of Hawking radiation, or more generally of quantum systems coupled to gravity. This was inspired in part by the Ryu-Takayanagi formula for calculating von Neumann entropy in holographic CFTs~\cite{Ryu_2006,Freedman2017}, and its generalisations~\cite{Hubeny2007,Engelhardt2014,Harper2018,Rolph2021}, and culminated in the island formula~\cite{Almheiri2019b} and quantum maximin prescription~\cite{Akers2019}. The island formula is a prescription for computing fine-grained entropy in quantum systems coupled to gravity in terms of semiclassical entropy. See~\cite{Almheiri2021} for a review of islands, including an explanation of what is meant by fine-grained, coarse-grained, renormalised and semiclassical entropy in this context. The island prescription is
\bne \label{eq:IslandFormula} S(R) = \min \text{ext}_I
 \left ( \frac{\text{Area}(\del I)}{4G_N} + S_{ren.}(R \cup I) \right ) + S_{c.t.} (R) \ene
where $S_{ren.}$ is the renormalised semiclassical von Neumann entropy, and $S_{c.t.}$ the entropy counterterm that isolates the UV divergences associated to the region's boundary. 
The prescription instructs us to extremise a functional over all possible islands $I$, including the trivial empty island $I = \varnothing$, then pick the minimal extremum.

The main technical obstacle to explicitly finding islands is evaluating $S_{ren.} (R\cup I)$ on the set of all possible islands. The cases where it is possible, such as in lower dimensional and doubly holographic models~\cite{Almheiri2019b,Almheiri2019,Almheiri2019a,Almheiri2020}, are sparse. This motivates finding necessary and/or sufficient conditions for the existence of islands. Sets of necessary and/or sufficient conditions for the existence of non-trivial islands may be easier to evaluate than the island formula~\eqref{eq:IslandFormula} itself. A set of necessary conditions were proposed in~\cite{Hartman2020}, and a set of sufficient conditions in the Island Finder paper~\cite{Bousso2021}. The literature which relies on these sets of conditions is growing, so it is important to know if and when they can be relied upon.

The two conditions given in~\cite{Bousso2021} for a region $R$ to have a non-empty island are in terms of an `island detector'\footnote{Credit to Ahmed Almheiri for the name island detector.} region $I'$ and require
\bne \label{eq:1aa} \qquad 1. \quad  S_{gen} (I' \cup R) < S_{gen} (R) \ene
and either
\bne  2.a. \label{eq:2aa}\quad \pm \Theta_\pm (I' \cup R) \geq 0 \ene
or
\bne 2.b. \label{eq:2bb}\quad  \pm \Theta_\pm (I' \cup R) \leq 0. \ene
where $\Theta_\pm$ is the quantum expansion in the future-directed outward $(+)$ and inward $(-)$ null directions normal to $\del I'$\footnote{See~\cite{Bousso2015} for a precise definition of quantum expansion.}.

In this paper we first point out a loophole in the proof that \cref{eq:1aa,eq:2aa,eq:2bb} are sufficient conditions for $R$ to have a non-empty island.
The proof given in~\cite{Bousso2021} is by contradiction: it is assumed that (1) there is an island detector region $I'$ satisfying conditions \eqref{eq:1aa} to \eqref{eq:2bb}, and (2) that the island $I$ is empty. Then it is shown that these two assumptions lead to a contradiction. The loophole is the implicit assumption that at least one of the time slices on which the empty island is the quantum maximin surface intersects the domain of dependence of $I'$\footnote{Assuming $I' \cup R$ is quantum normal, i.e.~\eqref{eq:2aa}. We will consider the quantum anti-normal case separately.}. If this assumption is \replaced{violated}{invalid} then the representative $\tilde{I}'$ - the intersection of $D(I')$ with the time slice - is itself empty and an intermediate step in the proof is invalid.  

We preempt and address counterarguments to this loophole which assert that the implicit assumption mentioned above is always valid. One counterargument is based on the claim that the past and future tips of $D(I')$ are truncated by singularities, and the basis of this claim are two puzzling contradictions that arise when considering small representatives $\tilde{I}'$ near either tip of $D(I')$.

The first puzzle is a direct tension with the quantum Bousso bound. Assuming that either of $\del D^\pm (I')$ is a quantum lightsheet, then the quantum Bousso bound implies that $S_{gen}(\tilde{I}' \cup R)$ must decrease as the representative $\tilde{I}'$ is deformed towards the tip of the $D^\pm (I')$. A contradiction, $S_{gen} (R) < S_{gen} (R)$, is found when the tip is reached and $\tilde{I}' = \varnothing$. We discuss several resolutions to this puzzle.
\deleted{(1) truncation of $D(I')$ by singularities, (2) $\del D^\pm (I')$ failing to be a quantum lightsheet due to caustics, and (3) to reach $\tilde {I}' = \varnothing$ one has to pass through UV regimes where the semiclassical relations like the quantum Bousso bound are invalid.}
\added{The first two are an evasion of the tension with the quantum Bousso bound: (1) there is no $\tilde{I}' = \varnothing$ because of truncation of $D(I')$ by curvature singularities, or (2) $\del D^\pm (I')$ fails to be a quantum lightsheet. The third resolution is that (3) the quantum Bousso bound \textit{is} violated, but that this is not unexpected because to reach $\tilde {I}' = \varnothing$ one has to pass outside the regime of validity of the semiclassical approximation.}
There is more than one resolution, and how the tension is resolved depends on the set-up.

The second puzzle is that the existence of island detectors naively and incorrectly implies that a ball-shaped region of the Minkowski vacuum violates the Bekenstein area bound. The authors of~\cite{Hartman2020} showed, with some mild assumptions, that any region $\mathcal{I}$ that satisfies $S_{gen.} (\mathcal{I} \cup R) < S_{gen.} (R)$, which includes island detector representatives, also violates the Bekenstein area bound. The second puzzle arises when applying this result to a representative that is smaller than the local curvature and temperature scales. If such a representative $\tilde{I}'$ exists for a given island detector $I'$, then the state on it is approximately that of the vacuum in flat space, which should not violate the Bekenstein area bound. 

\replaced{The puzzles are of broader importance than may initially appear, because they also need to be resolved for islands. As far as the puzzles are concerned islands are simply special cases of island detectors in the sense that they satisfy \eqref{eq:1aa} and saturate both conditions~\eqref{eq:2aa} and~\eqref{eq:2bb} by their definitional extremality.}
{The same puzzles need to be resolved for islands, even though they first arose when considering counterarguments to the loophole, because as far as the puzzles are concerned islands are simply special cases of island detectors in the sense that they satisfy \eqref{eq:1aa} and saturate both conditions~\eqref{eq:2aa} and~\eqref{eq:2bb} by their definitional extremality.}

We discuss in depth the resolution of these two puzzles for islands in asymptotically flat evaporating black holes. The past domain of dependence of the island has no curvature singularities to prevent one from considering arbitrarily small representatives\added{, so the quantum Bousso bound \textit{is} apparently violated}. A key part of resolving the Bekenstein bound puzzle is the sub-Planckian radial separation of the boundary of the island and the past lightcone of the radiation region\added{, which is an obstacle to specifying `nice' time slices with trace extrinsic curvature $|K| \ll l_p^{-1}$}.


\textbf{Outline}: In Sec.~\ref{sec:IslandFinder} we review Island Finder and explain the loophole in the proof. In Sec.~\ref{sec:Puzzles} we discuss counterarguments to the loophole, which naturally leads us to the two puzzles, to which we propose resolutions. In Sec.~\ref{sec:Discussion} we give some open questions unresolved by this work and ideas for future research. In Appendix~\ref{sec:Mirages} we outline two set-ups which may have explicit examples of island mirages. 

\section{Island Finder}\label{sec:IslandFinder}

In this section we discuss the Island Finder conditions and reproduce the proof given that they are sufficient conditions for the existence of a non-empty island~\cite{Bousso2021}. We point out loopholes in the proof. 

\subsection{Conditions}

\begin{figure}
     \centering
     \includegraphics[width=0.8\textwidth]{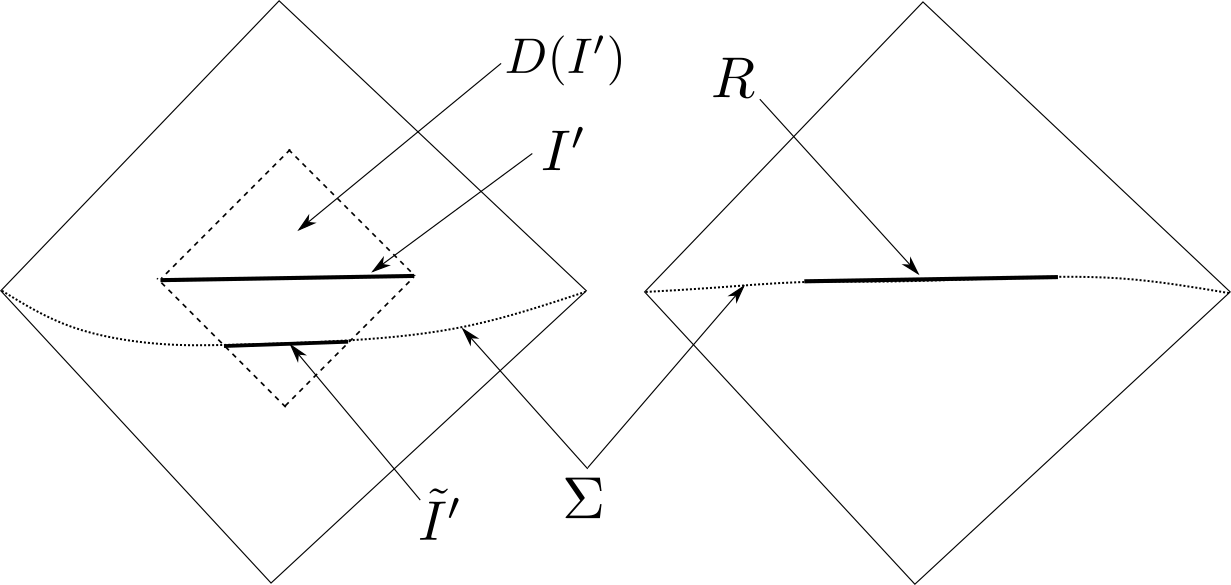}
     \caption{Two (asymptotically flat) quantum systems, one with dynamical gravity and one without. $R$ is a region in the non-gravitating system, and $I'$ a region in the gravitating system. $\Sigma$ is a time slice, and $\tilde{I}'$ the representative of $I'$ on that time slice as defined in~\eqref{eq:tildeI'def}. If $I'$ satisfies conditions \eqref{eq:cond1a} and \eqref{eq:cond2} then it is an island detector for $R$. \deleted{A loophole in the proof given in~\cite{Bousso2021} is the implicit assumption that $\Sigma$ intersects the domain of dependence $D(I')$.}}
     \label{fig:IslandProof}
\end{figure}


Consider two disjoint achronal regions, $I'$ and $R$, in a spacetime $M$ with $R$ in a weakly gravitating subregion. 
The claim of \cite{Bousso2021} is that if two conditions are satisfied by the generalised entropy of $I' \cup R$ then this is sufficient to guarantee that $R$ has a non-empty island somewhere in $M$. 

We will call an $I'$ that satisfies the two conditions an island detector region for $R$. If an island detector region exists and $R$ does have a non-empty island $I$, then the conditions are working as intended. If an island detector region exists and the island $I$ for $R$ \textit{is} empty, then we have a faulty island detector, and we call such an $I'$ an island mirage.

The first of the two conditions is
\bne \label{eq:cond1a} (1)\quad  S_{gen} (I' \cup R) < S_{gen}(R) \ene
There is an equivalent form of this inequality, 
\bne \label{eq:cond1b}
I(I',R) > S_{gen} (I') 
\ene 
which has the benefit of both sides being UV-finite even if we decouple gravity, $G_N \to 0$, in the region that contains $R$~\cite{Casini2007, Bombelli1986, Hartman2020}. One quick comment is $S_{gen}(I')$ scales as the surface area of $I'$ in Planck units, so there is a manifest need for a large degree of correlation between $I'$ and $R$ in order to satisfy the condition. 

The second of the two sufficient conditions is that $I'\cup R$ be either quantum anti-normal or quantum normal. Quantum normality is the property that the generalised entropy of a spatial region increases under deformations of the region's boundary in outward null directions. This is a generalisation of the classical notion of a normal spatial region whose surface area increases under outward deformations. The definition of quantum normality is 
\bne \label{eq:cond2}
(2) \quad \pm \Theta_\pm (I' \cup R) \geq 0
\ene
where $\Theta_\pm$ is the quantum expansion in the outward/inward future null directions~\cite{Bousso2015}. Quantum anti-normality is the opposite of quantum normality in the sense that the generalised entropy \textit{decreases} under outward null deformations. Any subregion of a time reflection symmetric slice is automatically quantum normal or anti-quantum normal, so any $I' \cup R$ on such a slice which satisfies condition \eqref{eq:cond1a} satisfies both of the Island Finder conditions. Also, if $I' \cup R$ is quantum normal then so is $I'$, which means the detector region $I'$ satisfies one of the necessary conditions to itself be an island~\cite{Almheiri2019a, Hartman2020}.  Note that \eqref{eq:cond2} is a property that is required to hold locally at every point on the boundary of $I' \cup R$, and that it depends on the non-local entanglement entropy.

\subsection{QFC and quantum maximin}

In the proof the quantum focussing conjecture (QFC) and quantum maximin prescription are assumed~\cite{Bousso2015, Akers2019}. We assume that the spacetime is stably causal, because that is the causality condition assumed for the quantum maximin prescription~\cite{Akers2019}, and because we do not want to exclude asymptotically AdS spacetimes\footnote{Classical maximin was originally introduced as a reformulation of the HRT holographic entanglement entropy prescription which applies to asymptotically locally AdS spacetimes~\cite{Hubeny2007, Wall2012}. Since AdS is not globally hyperbolic it has no Cauchy slices, so the notion of Cauchy slices had to be enlarged to AdS-Cauchy slices which are geodesically complete achronal surfaces~\cite{Wall2012, Engelhardt2014}}. Stably causal spacetimes have a global time function, and the function's level sets are constant time slices~\cite{Wald:1984rg}. 

Quantum maximin is a prescription for calculating entropy of subregions in non-holographic systems as well as holographic ones~\cite{Akers2019}. A relevant class of models are AdS black holes evaporating into a non-gravitating bath, where in the non-gravitational description we couple a holographic and non-holographic system together~\cite{Almheiri2019a}. If $R$ is a subregion of the non-holographic system, then the quantum maximin prescription for calculating its exact von Neumann entropy is
\bne \label{eq:qmaximin} S(R) = \max_\Sigma \min_{I\subset \Sigma} \left ( \frac{\text{Area}(\del I)}{4G_N} + S_{ren.}(R \cup I) \right )  \ene 

The quantum maximin prescription \eqref{eq:qmaximin} is equivalent to the island formula, assuming the QFC and a certain stability condition on $I$~\cite{Akers2019}. The prescription minimises over candidate islands on a given time slice, then maximises over time slices. Let us call such a maximising time slice $\Sigma$ a maximin slice. The maximin slice need not be unique, particularly if the island is empty.

\subsection{Proof}

The proof given in~\cite{Bousso2021} is a proof by contradiction. Start by assuming that the two conditions \eqref{eq:cond1a} and \eqref{eq:cond2} are satisfied, and that the true island $I$ is the empty set, i.e. there is no quantum extremal region $I$ that lowers the generalised entropy of $R$.  We will treat the case where $I' \cup R$ is quantum normal first, and consider the quantum anti-normal case separately. 

$I' \cup R$ is quantum normal so the quantum expansion in the inward null directions is initially negative. Consider a time slice $\Sigma$ that contains $R$ and define the representative $\tilde{I}'$ of $I'$ on $\Sigma$ as
\bne \label{eq:tildeI'def} \tilde{I}' := D(I') \cap \Sigma \ene
where $D(I')$ is the domain of dependence of $I'$.  QFC states that the quantum expansion cannot increase along a null congruence, so that the quantum expansion must remain negative along the null congruence from $I'$ to $\tilde{I}'$. The generalised entropy can only decrease under such inward null deformations:
\bne \label{eq:QFC_implication} \begin{split} S_{gen} (\tilde I ' \cup R) &\leq S_{gen} (I' \cup R)\\
& < S_{gen} (R)
\end{split} \ene
The second line follows from \eqref{eq:cond1a}. 

The contradiction appears if we take $\Sigma$ to be a time slice on which the codim-2 quantum maximin surface lives. 
From the quantum maximin prescription the true island $I$ gives the lowest generalised entropy on its maximin slice, and since we assumed $I$ to be the empty set we have that
\bne \begin{split} S_{gen}(\tilde I' \cup R) &\geq S_{gen} (I \cup R) \\ &= S_{gen} (R) \end{split} \ene
which directly contradicts \eqref{eq:QFC_implication}. The conclusion is that the island cannot be empty if the two conditions \eqref{eq:cond1a} and \eqref{eq:cond2} are satisfied.


When $I' \cup R$ is quantum anti-normal the essential steps in the proof are the same, except that the representative $\tilde{I}'$ of $I'$ on a time slice $\Sigma$ is defined as
\bne \tilde{I}' := J(I') \cap \Sigma \ene 
with $J(I')$ all points that can be reached from $I'$ with causal curves~\cite{Bousso2021}. 

\subsection{Loophole} \label{sec:loophole}
The proof in a nutshell is that the conditions imply that every time slice that contains $R$ and intersects $D(I')$ has a non-empty $\tilde{I}'$ for which $S_{gen} (\tilde{I}' \cup R) < S_{gen} (R)$. This rules out that time slice being the maximin slice for a true island that is empty, as the true island must give the lowest generalised entropy on the maximin slice. 

One loophole in the proof is the implicit assumption that the maximin slice of the empty island intersects $D(I')$. If the true island is empty, and its maximin slice does not intersect $D(I')$, then by the definition \eqref{eq:tildeI'def} we have $\tilde I' = \varnothing$, and the specific step that led to a contradiction that fails here is \eqref{eq:QFC_implication}. No contradiction is reached and the proof fails.

\deleted{The QFC does not constrain the generalised entropy of subregions of time slices that do not intersect $D(I')$.}
\added{The QFC does not constrain, in relation to $S_{gen}(I' \cup R)$, the generalised entropy of subregions $X\cup R$, with $X \neq \varnothing$, on time slices that do not intersect $D(I')$.}
QFC is defined on null congruences, and by definition the integral curves in a congruence are non-intersecting, so congruences end at caustics. QFC states that the quantum expansion cannot increase along a null congruence, but the tips of $D(I')$ are caustics. 
\added{The QFC \textit{does} constrain $S_{gen.}(X \cup R)$ for $X = \varnothing$ if we apply it to a quantum normal $I' \cup R$ and integrate it over the whole of either of $\del D^\pm (I')$:}
\bne \label{eq:contra} S_{gen.} (R) \leq S_{gen.} (I' \cup R) \ene
\added{Thus there is a direct tension between island detectors and the QFC, if there exists an empty representative $\tilde{I}' = \varnothing$, because~\eqref{eq:contra} contradicts~\eqref{eq:cond1a}. This is a loophole counterargument that we will return to in Sec.~\ref{sec:Quantum_Bousso_bound}.}

\deleted{This is related to another loophole in the proof, which is the implicit assumption that the null congruence to which the QFC is applied does not end before the past or future tips of $D(I')$ are reached. $\del D^{\pm}(I')$ will not always be a null congruence, because of the integral curves of null generators from $\del I'$ reaching caustics. Equation~\eqref{eq:QFC_implication} in the proof is not implied by QFC if $\tilde I'$ is on a time slice that intersects $D(I')$ but not the null congruence to which QFC is applied. }

\deleted{This second loophole is important, but it is difficult to make general statements about in which set-ups caustics will form and the null congruence end prematurely, so there is little more we can say without studying specific models. By contrast, in general set-ups there are time slices that do not intersect $D(I')$, so in what follows we will focus on the first, `empty representative' loophole.}

\begin{figure}
     \centering
     \includegraphics[width=0.65\textwidth]{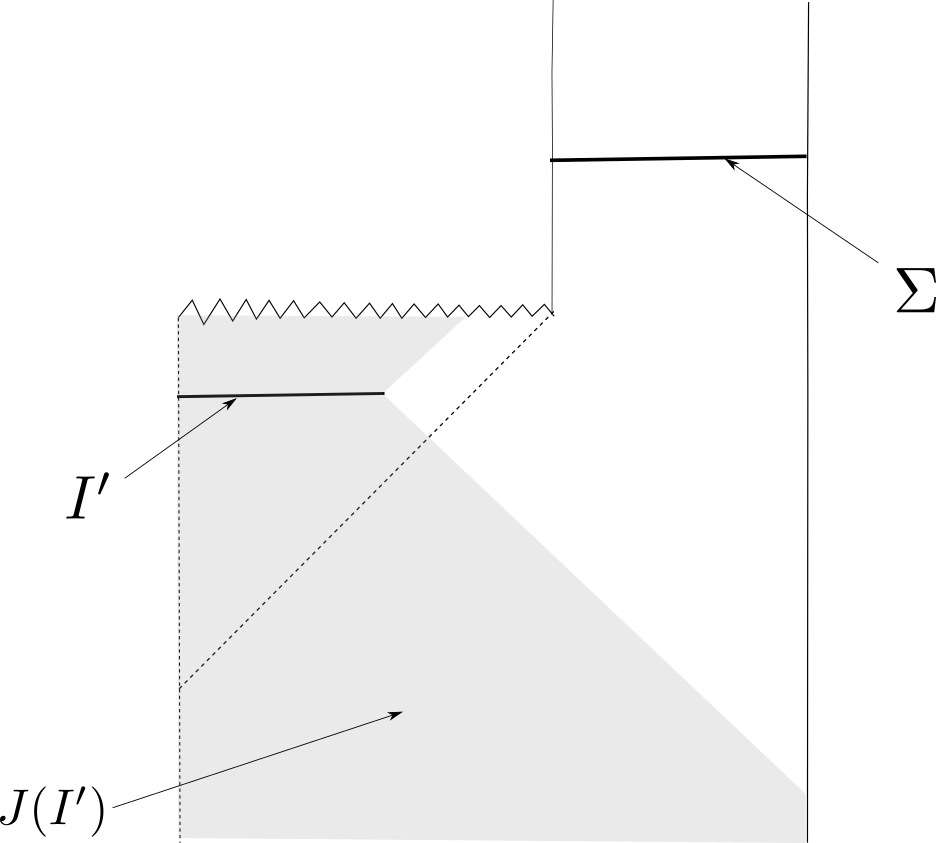}
     \caption{The Penrose diagram of a small AdS black hole that forms from collapse then evaporates away to nothing. $\Sigma$ is a time slice. $I'$ is a region whose representative $\tilde{I'} := \Sigma \cap J(I') = \varnothing$, for which the step~\eqref{eq:QFC_implication} of the proof in~\cite{Bousso2021} is incorrect. This is an example of the causal structure needed for $\tilde{I}'$ to be empty in the case of $I' \cup R$ being quantum anti-normal.}
     \label{fig:antiqnormal}
\end{figure}


When $I' \cup R$ is quantum normal any time slice that does not intersect $D(I')$ gives an empty representative $\tilde{I}'$. To have an empty representative when $I' \cup R$ is quantum anti-normal there must exist a time slice that does not intersect $J(I')$. This is not possible in globally hyperbolic spacetimes, because all causal curves from $I'$ will intersect all Cauchy slices. Recall however that quantum maximin is applicable not only to globally hyperbolic spacetimes, but also spacetimes that are stably causal and not globally hyperbolic. This allows for time slices on which $\tilde{I}'$ is empty. An example of a stably causal spacetime with an $I'$ and a time slice $\Sigma$ on which the representative $\tilde{I}'$ is empty is shown in Fig.~\ref{fig:antiqnormal}. Singularities are barriers to causal curves from $I'$. This figure is only an example of the causal structure needed for an island mirage when $I' \cup R$ is quantum anti-normal - we have not shown nor do we claim that the Island Finder conditions are satisfied in this case. 

Whether $I' \cup R$ is quantum normal or anti-normal, there are time slices on which the representative $\tilde I' = \varnothing$, and $S_{gen} (\tilde I' \cup R) \nleqslant S_{gen} (I' \cup R)$. No contradiction is reached, so a true island that is empty is not ruled out. The island finder conditions are insufficient to establish existence of islands, and the conditions may be satisfied without $R$ having a non-empty island. The conditions are sufficient only to establish nonexistence of empty islands whose maximin slice intersects the domain of dependence of $I'$ (when $I' \cup R$ is quantum normal). The concommitant upper entropy bound given in~\cite{Bousso2021} is also not proven if the proof that the conditions are sufficient for a non-empty island has a loophole.

\section{Loophole counterarguments and island puzzles} \label{sec:Puzzles}

\subsection{Time slices for empty maximin surfaces}

The first objection one may have to the loophole is that if the true island is empty, then is it not irrelevant which time slice we use when calculating the entropy? Can we not always choose a time slice with a non-empty intersection with $D(I')$? The loophole is indeed closed if all empty islands have maximin slices which intersect $D(I')$. As a reminder, we defined maximin slices to be the maximising subset of time slices in the quantum maximin prescription~\eqref{eq:qmaximin}.

Now it is true that when the island is empty the von Neumann entropy of $R \cup I$ is the same for all time slices that contain $R$, and that in general a subset of those time slices will intersect $D(I')$. What is not true is that all such time slices
are maximin slices for the empty island. If the true island is empty then by maximin there is certainly at least one time slice on which the empty-island entropy is the minimal entropy, but there is no guarantee that the no-island entropy is the minimal entropy on every time slice containing $R$.

Assuming that the island is empty, a time slice fails to be a maximin slice when on that slice there is a subregion $\mathcal{X}$  for which 
\bne \label{eq:noislandcond} \frac{\text{Area}(\del \mathcal{X})}{4G_N} + S_{ren.}(R \cup \mathcal{X})< S_{ren.} (R)\ene
As an example of this, consider a time slice that contains a subregion $\mathcal{X}$ whose boundary has zero area, i.e. whose components are null hypersurfaces, and for which $\mathcal{X}$ is entangled with $R$ such that $S_{ren.}(\mathcal{X} \cup R) < S_{ren.} (R)$. Similar to the setups considered in~\cite{Hartman2020}, we could take a thermofield double state on AdS and a non-gravitating bath, take $R$ to be the whole of the bath, and $\mathcal{X}$ to be a ball-shaped region in the AdS whose boundary has been deformed into zigzag union of many null hypersurfaces. The time slices that contain this $\mathcal{X}$ (and $R$) fail to be maximin slices for the empty island.
\cts{Think carefully about when/whether $S_{ren.}(AB) < S_{ren.}(A)$.}

We have argued that for an empty island not every time slice (containing $R$) is necessarily a maximin slice, but for the loophole to work we need something stronger: that no time slice that intersects $D(I')$ is a maximin slice when the island is empty. This requires that on all such time slices with $\Sigma \cap D(I') \neq \varnothing$ there exists an $\mathcal X$ with $\mathcal X \cup R \subseteq \Sigma$ 
such that~\eqref{eq:noislandcond} is satisfied.
Island Finder~\cite{Bousso2021} showed that~\eqref{eq:noislandcond} is satisfied by the representative $\tilde{I}'$ as defined in~\eqref{eq:tildeI'def}. Thus an $I'$ that satisfies conditions~\eqref{eq:cond1a} and~\eqref{eq:cond2} only rules out empty islands whose maximin slices intersect $D(I')$; it is not sufficient for the existence of a non-empty island.

\subsection{Quantum Bousso bound puzzle} \label{sec:Quantum_Bousso_bound}

Another objection to the loophole is the apparent contradiction with the quantum Bousso bound when there a time slice that does not intersect $D(I')$, with $I' \cup R$ satisfying conditions~\eqref{eq:cond1a} and~\eqref{eq:cond2}. 

The version of the quantum Bousso bound that we use is that which is implied by the QFC~\cite{Bousso2015}. There is a different version of the quantum Bousso bound that has been proven using positivity of relative entropy~\cite{Casini2008, Bousso2014}. In that version of the bound the entanglement entropy is regulated by vacuum subtraction, and vacuum-subtracted entanglement entropy is inequivalent to the renormalised entropy used in $S_{gen.}$~\cite{Bousso2014, Bousso2015}. We need to use the version implied by the QFC, because it is that version that constrains the renormalised and generalised entropies that appear in the island formula.

The quantum Bousso bound applies to quantum light sheets, which are null hypersurfaces on which the quantum expansion is every nonpositive. Since $I'\cup R$ is quantum normal by assumption, QFC implies that the past and future boundaries of the domain of dependence $D(I')$ are quantum light sheets, and the quantum Bousso bound states that~\cite{Strominger2003}
\bne \label{eq:qBoussoBound} S_{gen.}(I' \cup R) \geq S_{gen.} (\tilde I ' \cup R)
\ene
where as before, $\tilde{I}' := \Sigma \cap D(I')$. 

We assume that the spacetime is stably causal, which is the same causality condition assumed in the quantum maximin prescription, such that there is a global time function $t$. With a reparametrisation of $t$ we can take the $t=0$ slice to contain $I'$, and the $t=\pm 1$ slices to intersect $D(I')$ at the future and past tips of $D(I')$. \cts{Assuming a reparametrisation such that there is a level set that contains $I'$ seems stronger than stable causality.} In the limit of taking time slices to the tips of the domain of dependence, $|t| \to 1^-$, we have $\text{Area}(\del \tilde I') \to 0$. If we assume also that as we deform towards either tip of $D(I')$ that 
\bne \label{eq:dubious_limit} \lim_{|t| \to 1^-} S_{ren.}(\tilde{I}' \cup R ) = S_{ren.} (R) \ene
then the \added{$|t| \to 1^-$ limit} of the quantum Bousso bound~\eqref{eq:qBoussoBound} implies
\bne S_{gen.}(I' \cup R) \geq S_{gen.} (R), \label{eq:qbcont} \ene
which directly contradicts \eqref{eq:cond1a}. \added{Thus as we deform $\tilde {I}'$ towards either tip of $D(I')$ then there is a direct tension with the quantum Bousso bound even before the exact point of $\tilde{I}' = \varnothing$ is reached, if~\eqref{eq:dubious_limit} is true.}  

\begin{figure}
     \centering
     \includegraphics[width=0.7\textwidth]{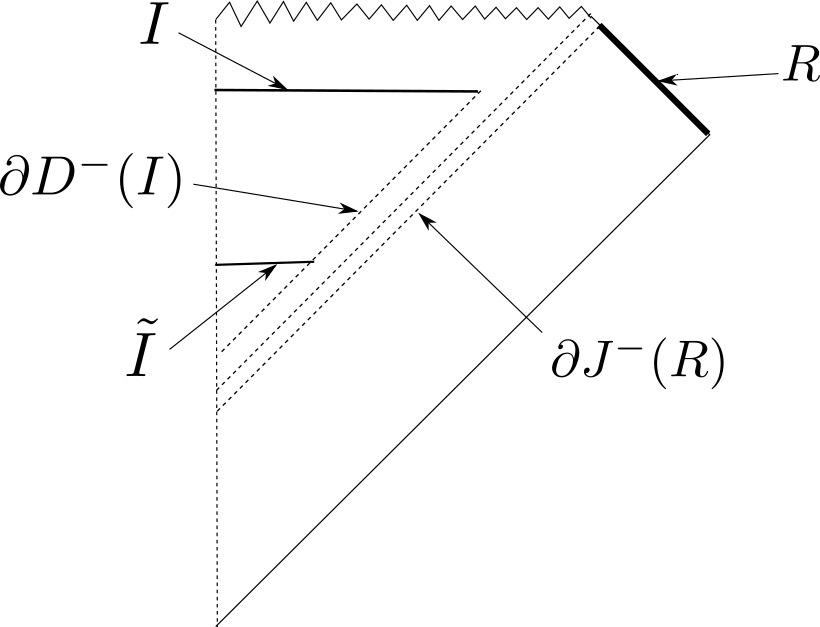}
     \caption{An evaporating black hole in asymptotically Minkowski spacetime. Radiation region $R$ has an island $I$. Also shown are the boundary of the causal past of $R$, the boundary of the past domain of dependence of $I$, and an earlier time representative $\tilde I$ of $I$. This is a counterexample to the claim that the existence of a region satisfying conditions~\eqref{eq:cond1a} and~\eqref{eq:cond2} implies the existence of singularities which remove both tips of the domain of dependence.}
     \label{fig:BHqboussobound}
\end{figure}

Islands, not just island detector regions, are in direct tension with the quantum Bousso bound by the same chain of logic. Recall that islands are special cases of island detector regions in the sense that they satisfy~\eqref{eq:cond1a} and saturate the bounds~\eqref{eq:cond2}. As an example, consider an evaporating black hole in asymptotically flat spacetime, as depicted in Fig.~\ref{fig:BHqboussobound}. The radiation region $R$ is taken to be sufficiently large to have an island $I$ in the black hole interior. The representative $\tilde{I} := \Sigma \cap D(I)$ on an earlier time slice satisfies
\bne \label{eq:35} S_{gen} (\tilde{I} \cup R) \leq S_{gen} (I \cup R) \ene
by the quantum Bousso bound and QFC. Again we reach the apparent contradiction $S_{gen} (R) < S_{gen} (R)$ when we reach the past tip of $D(I)$ and $\tilde{I} = \varnothing$. 

This contradiction needs to be resolved whenever conditions \eqref{eq:cond1a} and \eqref{eq:cond2} are satisfied, whether or not there is a non-empty island in the spacetime. Here are several ways the contradiction can be evaded:

\vspace{3mm}
1. $D(I')$ is truncated by curvature singularities.
\vspace{3mm}

One possible resolution is that when conditions~\eqref{eq:cond1a} and~\eqref{eq:cond2} are satisfied there are always singularities to the future and past of $I'$ that remove the tips of $D(I')$, so that the problematic limit~\eqref{eq:dubious_limit} does not exist. In some spacetimes this would force all time slices to intersect $D(I')$, so there would be no empty representatives $\tilde{I}'$ and the loophole would be closed
. If truncation by singularities were the only resolution to the tension with the quantum Bousso bound then conditions \eqref{eq:cond1a} and \eqref{eq:cond2} would also be sufficient conditions for the existence of singularities\footnote{This was also pointed out in~\cite{BoussoTalk}.}.





There is a simple counterexample to the claim that there must be singularities, which is the evaporating black hole of Fig.~\ref{fig:BHqboussobound}. While there is a singularity to the future of $I$ there is none to the past. Curvature singularities may resolve the tension with the quantum Bousso bound in some cases, but there must be a different resolution for the past domain of dependence of an island in an evaporating black hole.

\vspace{3mm}
2. $\del D^\pm (I')$ fails to be a quantum lightsheet\deleted{ due to caustics}.
\vspace{3mm}

Another way in which the quantum Bousso bound can fail to be applicable is if only a subset of $\del D(I')$ is a quantum lightsheet. Classical lightsheets, with everyhere negative classical expansion, can terminate due to the formation of caustics where $\theta \to -\infty$. Assuming the null energy condition is satisfied, the Raychaudhuri equation implies that caustics form, and so lightsheets \added{possibly} end after finite affine parameter whenever the classical expansion is negative~\cite{Raychaudhuri:1953yv}. The truncation of lightsheets underlied an argument against the critically illuminated black hole counterexample to the classical Bousso bound~\cite{Lowe1999}\footnote{Angular caustics also save the classical Bousso bound when there is anisotropy in a collapsing shell, which increases its entropy but also decreases the extent of the lightsheet~\cite{Bousso1999}. The critically illuminated black hole does not violate the vacuum-subtracted entropy version of the quantum Bousso bound~\cite{Bousso2014}. Vacuum-subtracted von Neumann entropy is not classical, in that it is not generally additive, so the resolution in~\cite{Bousso2014} is not applicable to the classical Bousso bound.}. 
\added{This argument against Lowe's counterexample was later realised to be flawed; $\del D(I')$ does \textit{not} fail to be a classical lightsheet, despite null geodesics in the congruence orthogonal to $\del I'$ meeting other null geodesics and caustics~\cite{Akers2018}}\footnote{Thanks to Raphael Bousso for discussions on this point.}.

\added{What is true for classical lightsheets is not necessarily true for quantum lightsheets. If quantum lightsheets can end after finite affine parameter, perhaps because of a quantum analogue of a classical caustic, $\Theta \to -\infty$, then the quantum Bousso bound can fail to apply once the null generator deforming $\del \tilde{I}'$ reaches that point.} 
\deleted{If it is true that quantum lightsheets end after finite affine parameter, because of a quantum analogue of a classical caustic, $\Theta \to -\infty$, then the quantum Bousso bound can fail to apply once the null generator deforming $\del \tilde{I}'$ reaches that point.} 
We will return to the question of whether $\Theta$ must diverge to $-\infty$ in finite affine parameter on a quantum lightsheet in the discussion section~\ref{sec:Discussion}. 


\deleted{It does not seem likely that caustics can resolve the tension with the quantum Bousso bound for islands for \textit{all} evaporating black hole set ups, though it may for critically illuminated black holes, as the formation of caustics is non-universal in that it depends on the stress-energy tensor of the infalling matter.}
\added{It does not seem likely that $\del D(I')$ failing to be a quantum lightsheet can resolve the tension with the quantum Bousso bound for islands for \textit{all} evaporating black hole set ups, as the truncation of quantum lightsheets seems non-universal in that it depends on the state, and in particular the stress-energy tensor of the infalling matter.}
We would like a more robust resolution for islands in evaporating black holes.


\vspace{3mm}
3. Blueshift effects.
\vspace{3mm}

There is a blueshift effect for Hawking modes near the horizon, and it is tempting to claim that the blueshift of Hawking modes towards the past to super-Planckian energies \replaced{is what leads to violations}{invalidates the use} of the \deleted{semiclassical} quantum Bousso bound\deleted{ and resolves the puzzle}, but we posit that this is not the case.

The renormalised entropy of the island comes from summing over interior Hawking modes below the renormalisation scale. As we propagate the interior modes backwards in time onto the representative region $\tilde I$ the relative blueshift - the ratio of frequencies as measured by free-falling observers that cross the event horizon at different times $v_1$ and $v_2$ - grows exponentially as~\cite{Jacobson2005}
\bne \frac{\omega_1}{\omega_2} \approx e^{(v_2 - v_1)/2r_s} \ene

Towards earlier infalling times the area of the boundary of the representative $\tilde I$ decreases, yet $\tilde I$ continues to be able to contain the same modes which purify the partner modes in $R$ despite the shrinking volume, due to the reduction in wavelength from the blueshift. Once the mode blueshifts to a frequency above the renormalisation scale, whatever that may be, it switches from contributing to $S_{ren.}$ to contributing to the entropy counterterm. Naively the blueshift effect leads to modes with super-Planckian energies that backreact on the geometry, a firewall. 

We are assuming no drama for infalling observers which implicitly fixes the entanglement between interior and exterior outgoing modes such that there is no backreaction~\cite{Almheiri2013}. Altering the state by removing a single Hawking mode from the island \textit{would} lead to a firewall in the past, which would  \replaced{lead to a curvature singularity}{invalidate the use of the quantum Bousso bound}.

To summarise, blueshift effects do not seem to resolve the puzzle because with the no drama assumption there are no curvature singularities in the neighbourhood of the past tip of $D(I)$ that could \replaced{explain the violation}{invalidate the use} of the quantum Bousso bound.

\vspace{3mm}
4. The naive result \eqref{eq:dubious_limit} was incorrect due to an unavoidable departure from the semiclassical regime when taking the limit. \added{Violations of the quantum Bousso bound are not necessarily unexpected when the bound is applied outside of the semiclassical regime, such as to regions of sub-Planckian size, or when it is derived from integrating the QFC outside of the semiclassical regime.}
\vspace{3mm}

The last resolution to the contradiction that we'll mention is that the naive result \eqref{eq:dubious_limit} is incorrect, because to take the limit you must consider regions $\tilde I'$ of Planckian size. Generalised entropy is a UV finite and regulator-independent quantity~\cite{Bianchi2012, Cooperman2013}, but it is a semiclassical quantity. The generalised entropy of a region of Planckian size is physically meaningless. It may be  mathematically well-defined if the semiclassical approximation is taken, in the sense that the region can at least be specified if quantum fluctuations of the background geometry are neglected, but the result has no relation to the physical reality. 

A simpler set-up where we can see what goes wrong if we trust formulas for entropy beyond their physical regime of validity is if we consider an interval of length $L$ in a 2d CFT in the vacuum state, whose entanglement entropy is $S(L) = \frac{c}{3} \log (L/\epsilon)$. We can evaluate $S(L)$ for $L< \epsilon$, but it physically meaningless and gives unphysical negative entropy. Just as in~\eqref{eq:dubious_limit}, we can take a naive limit 
\bne \lim_{L \to 0} S(L) = -\infty \ene
and get a result which is not the correct entropy of the empty set, which is zero. There is an important distinction between intervals of infinitesimal and zero length.

\deleted{The conclusion then is that as we shrink $\tilde{I}'$ towards the either tip of $D(\tilde{I}')$ it becomes so small that even if the generalised entropy of $\tilde{I}' \cup R$ can be formally evaluated it is physically meaningless. The semiclassical approximation is invalid and we are beyond the regime of applicability of the QFC. This resolution is how we believe the tension with the quantum Bousso bound is resolved for $\tilde I$ in the evaporating black hole set-up of Fig.~\ref{fig:BHqboussobound}.}

\added{The QFC and the quantum Bousso bound are not physically meaningful when applied to regions of sub-Planckian size, and violations of these bounds are not surprising in these applications as we are outside of the semiclassical regime.}

\added{
This does not immediately resolve the tension with the quantum Bousso bound when it is applied to the whole of $\del D^\pm$ as for an island detector in \eqref{eq:contra}, or for an island as in~\eqref{eq:35}, because in these cases no subregions of sub-Planckian size are directly involved in the bound. 



We may derive the quantum Bousso bound, as applied in \eqref{eq:contra}, by integrating the QFC over $\del D^\pm (I')$. To do so we must however apply the QFC to sub-Planckian regions, as we integrate towards either tip of $D(I')$, where violations of the QFC are not unexpected. A bound that can be derived by integrating the QFC along a null congruence like $\del D^\pm (I')$ is not always false - that would rule out the GSL~\cite{Bousso2015} - but violations of the bound are not necessarily unexpected since we are applying the QFC outside of the semiclassical regime.}   




\subsection{Bekenstein bound puzzle} \label{sec:BekBoundViol}
In the previous subsection we described a tension with the quantum Bousso bound that arises when taking \added{the} infinitesimal limit of the representative of either of an island or a quantum normal island detector\added{, as well as beyond that when the representative is empty}. In this subsection we describe a different puzzle that arises \textit{before} the representative reaches Planckian length scales. 

This puzzle is relevant to the loophole described in section~\ref{sec:loophole}, because just like the Bousso bound puzzle if the only resolution were that curvature singularities truncate $D(I')$ before the tension is reached then it is more difficult for the empty island maximin slice to not intersect $D(I')$, though it is still possible in a stably causal spacetime. 


Suppose there exists a region $R$ with island detector $I'$, with $I' \cup R$ quantum normal. As we follow null generators along $\del D(I')$, assuming for now that no \deleted{caustics or} curvature singularities are reached, then the QFC implies 
\bne \begin{split} \label{eq:starteq} S_{gen} (\tilde I' \cup R) &\leq S_{gen} (I' \cup R)\\ &< S_{gen} (R) \end{split} \ene
Now in~\cite{Hartman2020} it was argued that any $\tilde{I}'$ which satisfies \replaced{$S_{gen} (\tilde{I}' \cup R) < S_{gen} (R)$}{\eqref{eq:starteq}} also satisfies
\bne \label{eq:BekBekBound} S_{ren.} (\tilde{I}') > \frac{\text{Area}(\del \tilde{I}')}{4G_N} \ene
This means that the representative violates of the Bekenstein area bound. The puzzle arises when $\tilde{I}'$ is parametrically smaller than the local curvature and temperature scale, but still larger than $l_p$. When $\tilde{I}'$ is this small the reduced state on it is approximately the vacuum in flat spacetime, which should not violate the Bekenstein area bound. Since $\tilde I'$ is still larger than the Planck scale we can cannot claim that \replaced{violations of the QFC in~\eqref{eq:starteq} are expected due to the size of the $\tilde{I}'$.}{the QFC was invalid to apply}. \added{Let us assume that the QFC has not been violated and look for other resolutions.}

\cts{Need reference for this statement about the vacuum state?}    



This puzzle needs to be resolved for both islands and island detectors. Let us see how this puzzle manifests itself in a setup with islands. Take an asymptotically flat black hole formed from collapse, see Fig. \ref{fig:BHqboussobound}. By applying QFC to $S_{gen} (I \cup R)$ we get
\bne S_{gen} (\tilde I \cup R) \leq S_{gen} (I \cup R) < S_{gen.} (R) \ene
which again, by the argument given for condition 1 in~\cite{Hartman2020}, implies that the representative $\tilde I$ violates the Bekenstein area bound. If the state on $\tilde I$ is approximately the vacuum then it should not violate the Bekenstein area bound. 

How small $\tilde I$ needs to be before the state on it is approximately the vacuum depends on how the black hole was formed. For stellar collapse $\tilde I$ will be a ball-shaped region in the centre of the star soon after the event horizon formed, and it needs to be smaller than the local temperature scale. For a collapsing shell of radiation, the Vaidya metric, the geometry inside the shell of collapsing radiation is flat and $\tilde I$ can be just inside. In either case, QFC along with the island's quantum extremality property implies that the generalised entropy cannot increase as we deform along null generators from $I$ to $\tilde I$, and we again reach the incorrect conclusion that a ball-shaped region of the vacuum violates the Bekenstein area bound.

\subsubsection{Is Bekenstein's bound necessarily violated?}

To explore this puzzle further we will reproduce the derivation from~\cite{Hartman2020} that
\bne \label{eq:311} S_{gen.} (\mathcal{I} \cup R) < S_{gen.} (R) \implies S_{ren.} (\mathcal{I}) > \frac{\text{Area}(\del I)}{4G_N} \ene
We want to see what approximations were made in the derivation and if those approximations are ever invalid. In~\cite{Hartman2020} the region $\mathcal{I}$ was an island, while for us the region of interest is a representative $\tilde I'$ or $\tilde I$ of an island detector or an island, but since they all satisfy the left hand side of~\eqref{eq:311} the derivation is identical.



\begin{figure}
     \centering
     \includegraphics[width=0.65\textwidth]{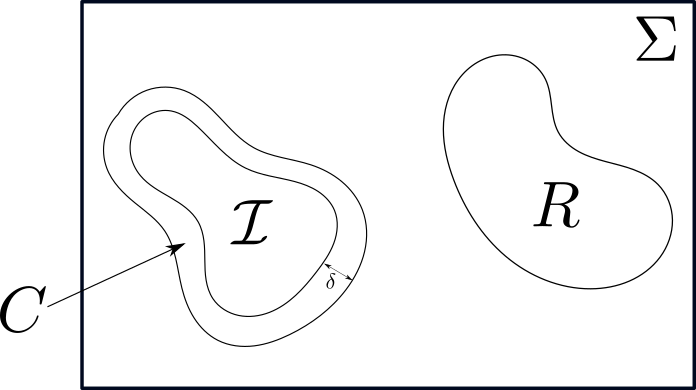}
     \caption{Representative cartoon of the subregions of time slice $\Sigma$ used in the proof of~\eqref{eq:311}.}
     \label{fig:BekBoundProof}
\end{figure}

The first step is to surround $\mathcal{I}$ with a thin region $C$ of width $\delta \gg l_p$. $C$ shares a boundary with $\mathcal{I}$, such that $\del \mathcal{I} \backslash \del C =\varnothing$, and is spacelike separated from $R$. See Fig.~\ref{fig:BekBoundProof}. Applying monotonicity of mutual information 
\bne I(X,Y\cup Z) \geq I(X,Y), \ene  
to the region complement to $\mathcal{I}\cup C$, which is a superset of $R$, we get 
\bne \begin{split} I(\mathcal{I},(\mathcal{I} \cup C)^c) &\geq I(\mathcal{I} ,R) \\
&> S_{gen} (\mathcal{I})
\end{split} \label{eq:BekBoundWithoutApproximations}\ene
We may rewrite \eqref{eq:BekBoundWithoutApproximations} in the following way, which makes its relation to the Bekenstein area bound transparent,
\bne \label{eq:BekBou} S_{ren.} (\mathcal{I}) - \frac{\text{Area}(\del \mathcal{I})}{4G_N} \geq S_{ren.} (\mathcal{I}) + S_{ren.} (C) - S_{ren.} (\mathcal{I}\cup C). \ene
This is done by expanding the mutual information in terms of the renormalised entropies,
\bne I(\mathcal{I},(\mathcal{I} \cup C)^c) = S_{ren.} (\mathcal{I}) + S_{ren.}(\mathcal{I} \cup C) - S_{ren.}(C), \ene
and rearranging. Note that no approximations have been made so far.

\cts{Is this step valid if $C$ has zero width? I think so as long as the boundaries of $C$ are finite.}

 We reach the conclusion that $\mathcal{I}$ violates the Bekenstein area bound at leading order in $G_N$, 
\bne S_{ren.} (\mathcal{I}) - \frac{\text{Area}(\del \mathcal{I})}{4G_N} \geq 0, \ene
if and only if the right hand side of \eqref{eq:BekBou} is non-negative or $O(G_N^0)$. Conversely, we need
\bne S_{ren.} (\mathcal{I}) + S_{ren.} (C) - S_{ren.} (\mathcal{I} \cup C) \lesssim - O(G_N^{-1}) \ene
in order to avoid the conclusion from \eqref{eq:BekBou} that $\mathcal{I}$ violates the Bekenstein bound. $\mathcal{I}$ and $C$ must have \textit{super}-additive renormalised entropies. Renormalised entropy can have unusual properties for an entropy, including superadditivity and negativity~\cite{Wall2009}, so this is not obviously impossible. The right hand side of~\eqref{eq:BekBou} is \textit{not} the mutual information between $\mathcal{I}$ and $C$, which would be non-negative by subadditivity, because the overlap of the boundaries $\mathcal{I}$ and $C$ is nonzero so there is a leftover piece when the boundary counterterms are added back in.


In~\cite{Hartman2020} it is assumed that 
\bne \label{eq:ass1} S_{ren.} (\mathcal{I} \cup C) = S_{ren.}(\mathcal{I}) \ene
and
\bne \label{eq:ass2} S_{ren.} (C) = 0 \ene
up to corrections that are subleading in $G_N^{-1}$ with respect to the area term. This is a reasonable assumption if there exists a time slice on which the state is semiclassical and with a subregion $C$ as defined before, and of width $\delta \gg l_p$ but smaller than the local energy and curvature scales. Then $S_{ren.} (C) \sim \delta^{-(d-2)} \ll G_N^{-1}$. As a first comment, note that one class of examples where there is no $C$ of width $\delta \gg l_p$ is if $\del \mathcal{I}$ and $\del R$ are null-separated. 

\begin{figure}
     \centering
     \includegraphics[width=0.5\textwidth]{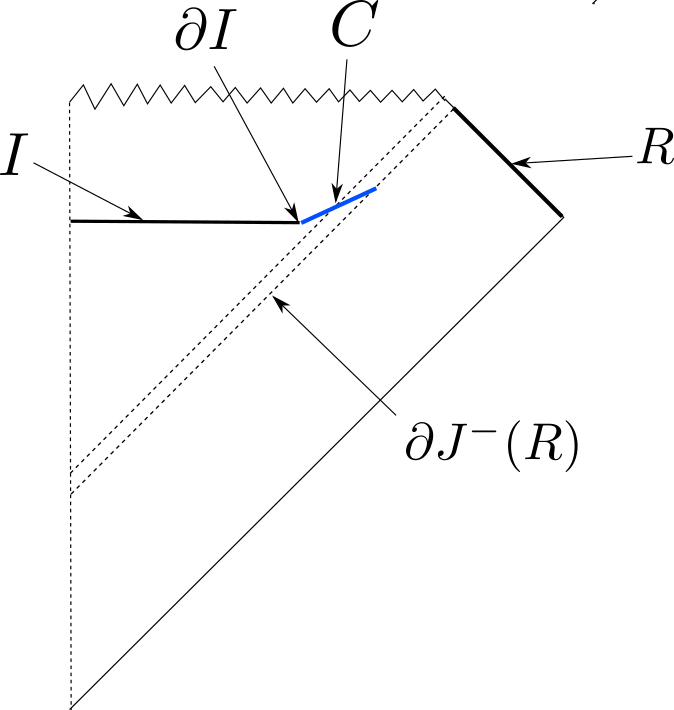}
     \caption{An evaporating black hole in asymptotically Minkowski spacetime. $C$ surrounds the island $I$, with $\del I \backslash \del C = \varnothing$, and is spacelike separated from $R$. We show that $C$ cannot have both width $\delta \gg l_p$ and trace extrinsic curvature $|K| \ll l_p^{-1}$.}
     \label{fig:CFigure}
\end{figure}

\subsubsection{Resolution for islands in evaporating black holes}

As a first step to resolving this subsection's Bekenstein bound related puzzle for an island in an asymptotically flat evaporating black hole, let us re-examine the assumptions~\eqref{eq:ass1} and~\eqref{eq:ass2} for this case. We will show that there is no time slice on which (1) there exists a $C$ of proper width $\delta \gg l_p$ and (2) the curvature scale of the slice, as quantified by its trace extrinsic curvature $K$, is not Planckian. 

The boundaries of $R$ and the island are not null-separated as can be seen in Fig.~\ref{fig:CFigure}, so there \textit{are} time slices with regions $C$ of non-zero width. We are working in the vacuum with respect to an infalling observer, so a natural set of coordinates to use are Painlev\'e-Gullstrand:
\bne ds^2 = - dT^2 + \left (dr + \sqrt{\frac{r_s}{r}}dT \right)^2 + r^2 d\Omega_{d-2}^2 \ene
where $T$ is the proper time for a radially infalling observer initially at rest at infinity. We will do an analysis in a region  perturbatively smaller than the length scale set by the black hole evaporation time, so we neglect the $T$ dependence of $r_s$.

The boundaries of $R$ and $I$ are not null-separated, but on a constant infalling time slice the radial separation of the island and the past lightcone of $R$ is Planckian~\cite{Penington2019}, so on that particular time slice there is no $C$ of spatial width $\delta \gg l_p$.
Let $r_{I}$ be the radial position of the boundary of the island on a constant $T$ slice, and $r_{lc}$ the radial position of the intersection of $\del J^- (R)$ with that slice. The reason why the radial separation $(r_{lc} - r_{I})$ must be Planckian is because the gradient of the entropy must be order $O(1/G_N)$ in order to compete with the $O(1)$ gradient of area in the extremality condition
\bne \label{eq:extrem_cond} \del_{r_I} \left ( \frac{\text{Area}(\del I)}{4G_N} + S_{ren.} (I\cup R) \right ) = 0 \ene
The gradient of the area is $\del_{r_I} \text{Area}(\del I) = 8\pi r_I$, while by dimensional analysis the leading divergence of the renormalised entropy is $\del_{r_I} S_{ren.} (I \cup R) \sim 1/(r_{lc} - r_I)$~\cite{Bousso2021}.
To satisfy the extremality condition \eqref{eq:extrem_cond} thus requires a radial separation $(r_{lc} - r_I) \sim l_p^2 / r_I$. 

The representative $\tilde I$ at earlier times has an even smaller radial separation from $\del J^- (R)$ than $I$, as the separation of lightrays from the event horizon decreases exponentially in advanced time $v := t + r^*$ towards the past. The maximum width an annular region $C$ surrounding $\tilde I$ on a constant $T$ slice can be is exponentially smaller than for $I$.

Now we discuss other time slicings, as 
we have only shown that there is no $C$ of width $\delta \gg l_p$ on the constant $T$ slice that intersects $\del I$. There are certainly other coordinate systems with time slices on which there is a $C$ of arbitrarily large spatial width, because the null separation of $\del R$ and $\del I$ in the future outward direction is infinite\footnote{It may seem counterintuitive that two nearly-null separated points can have large spatial separation. To get an intuitition for this, consider the spatial geodesic connecting two spacelike separated points in 2d Minkowski. This has proper length $-\Delta x^+ \Delta x^-$ which diverges when $\Delta x^+ \to \infty$ if $\Delta x^-$ is fixed and finite.}. 

$C$ is annular with inner boundary fixed to be $\del I$. For a fixed time slice $\Sigma$ the outer boundary of the maximum width $C$ is $\Sigma \cap \del J^- (R)$, which is as far as $C$ can extend radially while staying spacelike separated from $R$, as shown in Fig.~\ref{fig:CFigure}. 

For fixed $I$ we can increase the maximum width $C$ by deforming the time slice such that $\Sigma \cap \del J^- (R)$ moves towards future null infinity. The cost of increasing the maximum possible width of $C$ by deforming the time slice between $\del I$ and $\del J^- (R)$ towards the null limit is that the slice in this region approaches an annulus-like subregion of a null cone, and null cones have a divergent trace extrinsic curvature.

Let us assume that $C$ stays in the neighbourhood ($\Delta T , \Delta r \ll r_s$) of $\del I$, and justify that assumption later. The geometry in this neighbourhood is approximately
\bne \label{eq:approx_geom} ds^2 \approx - dT^2 + d\rho^2 + r_s^2 d\Omega^2_{d-2} \ene
where $\rho$ is the Lema\^itre radial coordinate, which is related to the Schwarzschild radial coordinate by
\bne r(T , \rho) = \left( \frac{3}{2}(\rho - T) \right )^{2/3} r_s^{1/3}. \ene
The radial separations of $\del I$ and $\del J^- (R)$ on a constant $T$ slice are approximately the same in the two coordinate systems
\bne (\rho_{lc} - \rho_i) \approx (r_{lc} - r_i) \sim l_p^2 / r_s \ene
because near $r = r_s$ we have $d\rho \approx dr + dT$.

Now consider a spherically symmetric time slice that intersects $\del I$ and in its neighbourhood has slope $d\rho / dT = \sqrt{1+\alpha^2}$ with $\alpha > 0$ to keep the slice spacelike. The trace extrinsic curvature of this time slice in the neighbourhood of $\del I$ is the same as the surface of a cone in Minkowski spacetime, which is 
\bne \label{eq:Kform} |K| = \frac{1}{\alpha r_s}. \ene
This $K$ diverges in the null limit $\alpha \to 0^+$. Recall that to maximise the possible width of $C$ we want to move the intersection $\Sigma \cap \del J^- (R)$ towards future null infinity, so let us take $1 \gg \alpha \gg l_p / r_s$, which is as close to the null limit as possible while keeping the extrinsic curvature sub-Planckian, $|K| \ll l_p^{-1}$. The slice intersects $\del J^- (R)$ after time 
\bne (\sqrt{1+\alpha^2}-1)\Delta T = (\rho_{lc} - \rho_{I}) \sim \frac{l_p^2}{r_s} \ene
Thus $\Delta T \ll r_s$ given the range of $\alpha$, which justifies the earlier \replaced{approximation~\eqref{eq:approx_geom}}{assumption that the geometry $C$ is embedded in is approximately flat}. The largest possible $C$ on the time slice defined has width given by the radial geodesic length between $\del I$ and the intersection of the slice with $\del J^- (R)$, which is
\bne  \delta = \alpha \Delta T \ll l_p . \ene

The conclusion is that there is no time slice on which there exists a $C$ with both width $\delta \gg l_p$ and with $|K| \ll l_p^{-1}$. 

We have not shown that islands satisfy the Bekenstein bound, nor do we claim that they do. We have only shown that one particular assumption that is sufficient to show that islands necessarily violate the bound is invalid for the black hole set up of Fig.~\ref{fig:CFigure}. The expectation is that islands violate the Bekenstein bound, while their early time representatives $\tilde{I}$ of size below the local curvature scale, as depicted in Fig.~\ref{fig:BHqboussobound}, do not. In order to avoid the conclusion that the Bekenstein bound is violated we have seen from equation~\eqref{eq:BekBou} that it is necessary for all possible regions $C$ to give strongly ($\sim G_N^{-1}$) superadditive renormalised entropy. 

Let us make some preliminary remarks on how $\tilde I$ and $C$ can have superadditive renormalised entropy, but leave a careful analysis for future work. We have shown that the largest width $C$ can have on a time slice with $|K| \ll l_p^{-1}$ is sub-Planckian for $I$, and it will be exponentially smaller still for $\tilde I$ because of the convergence towards the event horizon of past-directed ingoing light rays. As mentioned in Sec.~\ref{sec:Quantum_Bousso_bound}, semiclassical entropies of regions of sub-Planckian size can be formally evaluated, even if they are not physically meaningful. If $S_{ren.} (\tilde I \cup C) \approx S_{ren.} (\tilde I)$, which seems reasonable as $L_{\tilde I} \gg l_p \gg L_{C}$, then superadditivity requires $S_{ren.}(C) < 0 $. Regions of width smaller than the renormalisation scale can have negative $S_{ren.}$. For example the renormalised vacuum entropy of an interval in a 2d CFT is proportional to $\log (L/\mu)$, with $\mu$ the renormalisation scale, which is negative for $L < \mu$.   For $\tilde I$ the maximum width $C$ can be is exponentially smaller than $l_p$, so $S_{ren.}(C) \lesssim -O(G_N^{-1})$ may be possible.

\section{Discussion and future work}\label{sec:Discussion}

In section~\ref{sec:loophole} we pointed out a loophole in the proof given in~\cite{Bousso2021} that their set of conditions are sufficient for the existence of a non-empty island. There are trivial fixes to their set of conditions to make them sufficient, such as an additional condition that all time slices in the spacetime intersect $D(I')$ (or $J(I')$ if $I' \cup R$ is anti-quantum normal rather than quantum normal). This particular additional condition would be undesirably restrictive though: it is satisfied for the recollapsing FRW universe example of~\cite{Hartman2020} but not for evaporating asymptotically AdS or flat Schwarzschild black holes. The bottom line is that we do not have a set of conditions that are (1) sufficient to establish existence of a non-empty island, (2) easier to evaluate than the island formula, and (3) are not overly restrictive in that they fail to detect islands for whole classes of examples where islands are expected or have been explicitly found.

Having pointed out loopholes in the proof, it is natural to look for explicit counterexamples, i.e. island mirages. In appendix~\ref{sec:Mirages} we outline two promising setups which we intend to explore further. There may be examples of island mirages which are very different in character from those suggested as possibilities in the appendix.

In section~\ref{sec:Quantum_Bousso_bound} we mentioned that it is unclear when or whether a quantum lightsheet will truncate after finite affine parameter. Such a result would be important as it limits the applicability of the quantum Bousso bound. Classically, \replaced{null geodesics in a congruence}{lightsheets} with negative expansion \replaced{reach}{end at} caustics after finite affine parameter because of the bound
\bne \label{eq:classfocussing} \theta' \leq -\frac{1}{d-2}\theta^2 \ene
which follows from the Raychaudhuri equation for null geodesic congruences and assuming the null energy condition. Similarly, a negative quantum expansion leads to the `quantum caustic' $\Theta \to -\infty$ after finite affine parameter if the following quantum analogue of~\eqref{eq:classfocussing} is true:
\bne \label{eq:strong-QFC} \frac{\delta}{\delta V(y_2)} \Theta[V;y_1]\leq -\frac{1}{d-2} \Theta (y_1) \Theta (y_2) \delta (y_1 - y_2) \ene
using the notation from~\cite{Bousso2015}.  This conjecture~\eqref{eq:strong-QFC} is stronger than QFC. The on-diagonal components give \eqref{eq:classfocussing} in the classical limit, and give a stronger version of QNEC in the weak gravity limit\footnote{Which in 4d may not be satisfied by a ball-shaped region of the CFT vacuum in Minkowski spacetime. We thank Aron Wall for pointing this out.}. We feel that these conjectures and their implications for the quantum Bousso bound are worth further study.

In section~\ref{sec:BekBoundViol} we described and gave possible resolutions to the Bekenstein bound puzzle, and discussed the puzzle in the context of islands in evaporating black holes in detail. We did not however give a complete argument for how the early time representative $\tilde I$ of the island can avoid the implication that it violates the Bekenstein bound with superadditive renormalised entropies, as it seems it must. For future work we believe that the islands in AdS$_2$ black holes found in~\cite{Almheiri2019} would be a fruitful arena in which to explore this puzzle in explicit calculational detail.

\acknowledgments

I would like to thank Tarek Anous, Raphael Bousso, Sean Colin-Ellerin, Ricardo Espindola, Jackson Fliss, Bahman Najian, Dominik Neuenfeld, Dora Nikolakopoulou, Edgar Shaghoulian, Arvin Shahbazi-Moghaddam, Shan-Ming Ruan, Manus Visser, and Aron Wall for useful discussions. Special thanks to Ben Freivogel for many productive discussions and comments on an earlier draft. This research was supported by the Stichting Nederlandse Wetenschappelijk Onderzoek Instituten (NWO-I). 
\appendix
\section{Possible island mirages} \label{sec:Mirages}


An island mirage is an island detector $I'$ that satisfies conditions~\eqref{eq:cond1a} and~\eqref{eq:cond2}, but for which $R$ has no island. Mirages, if they exist, are counterexamples to the claim that the conditions are sufficient to guarantee that $R$ has an non-empty island somewhere. In this appendix we outline two setups which we were not able to explicitly show have island mirages, but that we feel are candidates worth further study in future work. 

\subsection{Small, critically illuminated, isolated AdS black holes}

\begin{figure}
     \centering
     \includegraphics[width=0.8\textwidth]{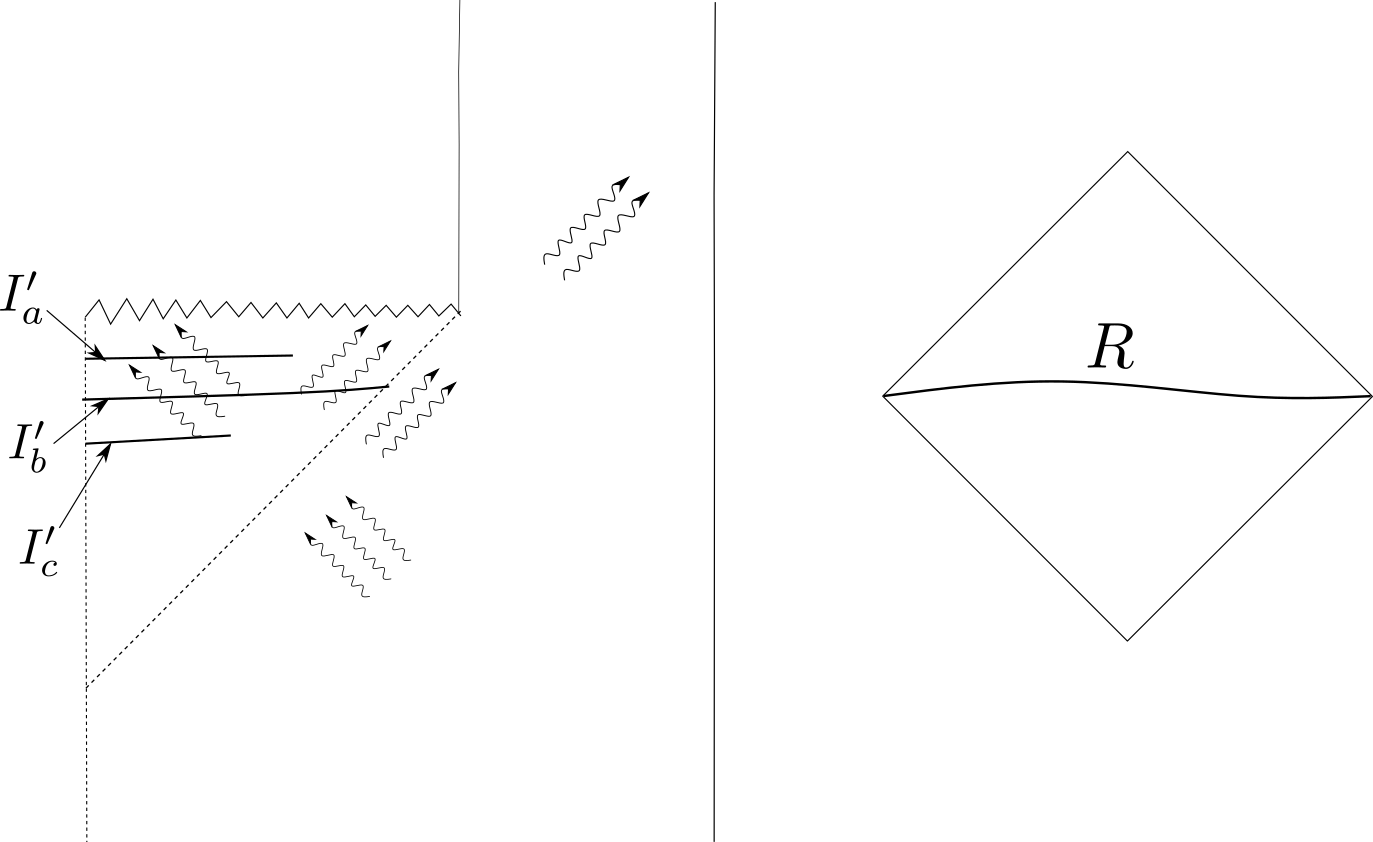}
     \caption{A small, critically illuminated, isolated AdS black hole, and an auxiliary non-gravitating flat spacetime. The radiation shown includes the incoming radiation that keeps the black mass constant, the outgoing Hawking modes, and their interior partners. The ingoing radiation is maximally mixed and purified by the degrees of freedom in $R$ in an auxiliary spacetime.}
     \label{fig:CriticallyIlluminated}
\end{figure}

Here we consider a small isolated AdS black hole that is critically illuminated by mixed state radiation that is purified by an auxiliary non-gravitational system, see Fig.~\ref{fig:CriticallyIlluminated}. Unlike typical island setups, isolated AdS black holes are not coupled to a non-gravitational bath; the asymptotic boundary conditions are reflecting for all time. Critically illuminated black holes have an incoming flux of energy that exactly balances the outgoing flux due to Hawking radiation, such that black hole has constant mass for some finite period of time~\cite{Lowe1999}. In our setup what an external observer sees is the formation of a small AdS black hole from collapse, that is then critically illuminated for some Schwarzschild time $t_{illum.}$, then allowed to evaporate away before the earliest Hawking radiation has had time to reflect and return from the boundary. For AdS black holes to be small enough to evaporate requires a mass less than
\bne \label{eq:massbound} M l_{AdS} \ll \left ( \frac{l_{AdS}}{l_p} \right )^\frac{d^2 -1}{2d -1} \ene
and there are no small AdS$_d$ black holes in $d \leq 3$~\cite{Harlow2016}. 

The parametric separation of time scales is taken to be
\bne t_{p} \ll t_{evap.} \ll t_{illum.} \ll l_{AdS} \ene 
There are several reasons to look for an island mirage in such a setup:

\begin{enumerate}
\item Isolated AdS black holes do not have islands. Every point in the black hole interior lies on a past causal horizon, so GSL rules out any quantum extremal surface~\cite{Jacobson2003, Engelhardt2021}.
\item Critically illuminated black holes are a counterexample to the classical Bousso bound as an arbitrarily large amount of classical entropy can pass through the event horizon while the horizon area stays constant~\cite{Lowe1999}. An arbitrarily large amount of entropy passing through the horizon is helpful for satisfying condition~\eqref{eq:cond1a}.
\item Small isolated AdS black holes only differ from asymptotically flat black holes, which do have islands, far from the black hole.
\end{enumerate}

Let us give a few comments about critically illuminated black holes. They are a counterexample to the classical Bousso bound, though not to GSL as the Hawking radiation continually adds to the entropy outside the black hole but it does not cross the lightsheet~\cite{Lowe1999}. The classical Bousso bound is in effect violated by quantum effects in the form of Hawking radiation. The critically illuminated black hole requires fine tuning of the incoming radiation, and microscopic fluctuations in the incoming radiation generally leads to the lightsheet truncating in finite affine parameter due to caustics
~\cite{Flanagan2000, Bousso2000, Bousso2002}. 

Since there can be no islands in isolated AdS black holes, to find an island mirage we just need to find an $I'$ that satisfies the conditions \eqref{eq:cond1a} and \eqref{eq:cond2}. In Fig.~\ref{fig:CriticallyIlluminated} we depict three such candidate $I'$ wich we will now discuss in turn. $I'_a$ satisfies \eqref{eq:cond1a} because it contains all the ingoing radiation so $S_{ren.}(I'_a \cup R) \approx 0$, and the boundary of $\del I'_a$ has small area, but it fails to be quantum normal because it is deep in the classically trapped region. $I'_b \cup R$ is quantum normal by GSL, because $\del I_b'$ lies on both a future and past causal horizon, but it seems likely to fail to satisfy \eqref{eq:cond1a} because it contains all the interior modes. 
$I'_c$ is in a sense in between $I'_a$ and $I'_b$, the Goldilocks candidate. The future inward null quantum expansion is non-positive by GSL, but since $\del I'_c$ is inside the event horizon it does not lies on a future causal horizon, so GSL does not apply to the future outward direction. To be quantum normal it must be outside the quantum trapped region, yet it must also contain enough of the incoming radiation to satisfy~\eqref{eq:cond1a}, and it is not clear whether both of these can be simultaneously satisfied.

\subsection{Large AdS black hole with infalling shell}
\begin{figure}
     \centering
     \includegraphics[width=0.99\textwidth]{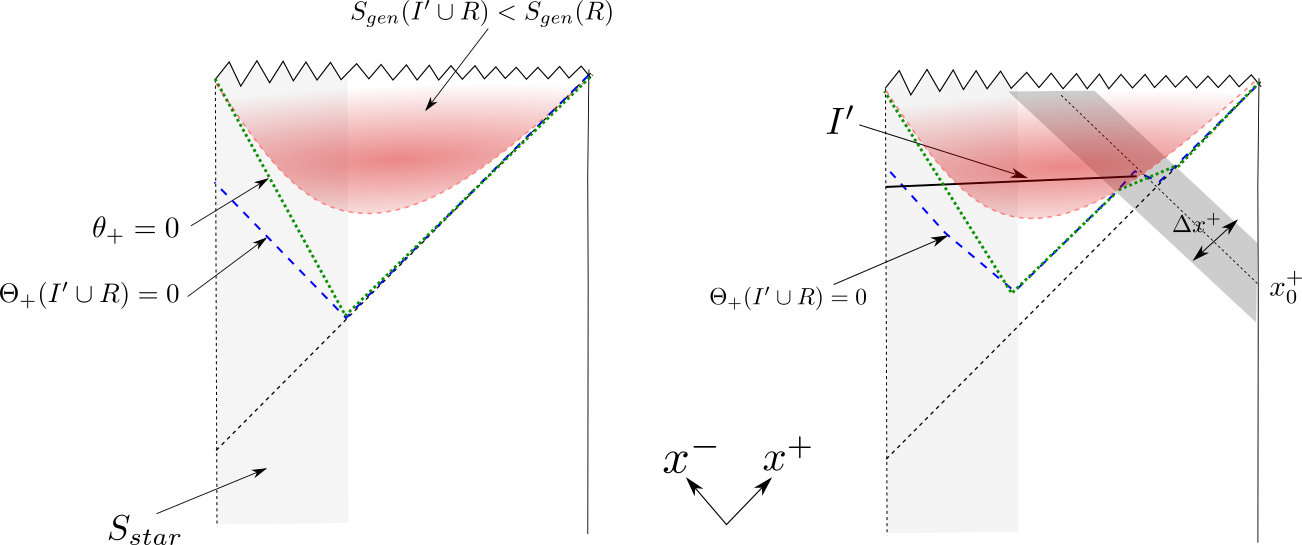}
     \caption{Left: A large, isolated AdS black hole formed from a collapsing star. The star is in a mixed state of entropy $S_{star}$ that is purified by $R$ in an auxiliary non-gravitating spacetime. Spherical symmetry is assumed, and $I'$ are ball regions whose boundaries are points in the figure. To be an island mirage $I' \cup R$ must be both outside the quantum apparent horizon, depicted in blue, and the inside the red hyperentropic region. Right: A null shell of pure radiation is added. The null shell has a large entropy gradient in the $x^+$ direction, so as to push the quantum apparent horizon into the red hyperentropic region. }
     \label{fig:NullShell}
\end{figure}

In this example we consider a large, isolated AdS black hole formed from a collapsing star, as shown in Fig. \ref{fig:NullShell}. The star is in a mixed state, with entropy $S_{star}$. The star is purified by $R$ in an auxiliary non-gravitating spacetime.
As an isolated AdS black hole, GSL again rules out non-trivial quantum extremal surfaces, so there are no islands. We assume spherical symmetry and look for ball-shaped island mirages $I'$, centred on the origin, which satisfy the conditions \eqref{eq:cond1a} and \eqref{eq:cond2}. 

To satisfy \eqref{eq:cond1a} and lower the generalised entropy, the boundary of $I'$ must be in the red region of Fig. \ref{fig:NullShell}. There the surface area of $I'$ is small and $I'$ contains a sufficiently large fraction of the collapsing star, which by construction is entangled with $R$, to lower the generalised entropy. To satisfy \eqref{eq:cond2}, $\del I'$ must be outside the quantum trapped region, whose boundary is the quantum apparent horizon depicted in blue in Fig.~\ref{fig:NullShell}. 

Take a point at the radial origin of the star after it has collapsed inside its Schwarzschild radius. The subset of the future directed null cone that starts from this point to a cut of the lightcone outside the classical trapped region is a lightsheet. If we assume the star's entropy to be extensive, then the classical Bousso bound places an upper limit on the entropy flux through the lightsheet
\bne S(L) \leq \frac{\text{Area}(\del L)}{4G_N}. \ene
\cts{$<$ or $\leq$?} This rules out an $I'$ whose boundary is outside the classically trapped region, and whose classical entropy violates the Bekenstein bound. If we assume that this island detector region violates the Bekenstein bound, i.e. there is a $C$ for which $I'$ and $C$ do not have superadditive renormalised entropies (see section~\ref{sec:BekBoundViol}), then we reach the conclusion that an island detector cannot exist outside the classically trapped region if the entropy of the matter forming the black hole is approximately classical, i.e. extensive. 
\cts{Think carefully about that statement.}

We need to add matter whose entropy is not classical in order to evade the classical Bousso bound, in order to have an $I'$ that is both quantum normal and violates the Bekenstein bound~\cite{Hartman2020}. We would like to `push' the quantum apparent horizon inward, as depicted in the right hand side of Fig. \ref{fig:NullShell} by creating a large positive entropy gradient in the $x^+$ direction, so making $\Theta_+$ positive where it was negative before. 


We can try adding a thin shell of radiation of thickness $\Delta x^+$. Take the shell to be composed of two concentric layers that are maximally entangled with each other, but with the whole shell in a pure state. A ball shaped region that is concentric with the shell will then have a large entropy gradient as a function of radius at the shell radius. 
A positive entropy gradient in the $x^+$ direction has the potential to flip the sign of $\Theta_+$ and so push the quantum apparent horizon in. 

We do have to be careful about the backreaction of the null shell  though and the effect on the classical expansion. One promising sign is that the quantum apparent horizon of the right hand side of Fig. \ref{fig:NullShell} seems not to violate the QFC. The QFC would forbid $\Theta_+ (I' \cup R)$ from changing signs twice as we deform $I'$ in the $x^+$ direction, but here it only changes sign once. A more careful analysis of the interplay between the entropy gradient and backreaction of the null shell is needed before this is a convincing example of an island mirage.

\cts{Add that radiation is free (pressure would stop it collapsing)}

\bibliography{references}
\bibliographystyle{JHEP}

\end{document}